# Social Innovation and the Evolution of Creative, Sustainable Worldviews

Liane Gabora and Mike Unrau

Department of Psychology, University of British Columbia, Kelowna BC, Canada

Running Head: SOCIAL INNOVATION EVOLUTION WORLDVIEWS

**Abstract**

Ideas build on one another in a manner that is cumulative and adaptive, forming open-ended lineages across space and time. Thus, human culture evolves. The pervasiveness of cross-domain creativity—as when a song inspires a painting—would appear indicative of discontinuities in cultural lineages. However, if what evolves through culture is not discrete 'memes' or artifacts but worldviews—the webs of thoughts, ideas, and attitudes that constitute our ways of seeing and being in the world—then the problem of discontinuities is solved. The state of a worldview can be affected by information assimilated in one domain, and this change-of-state can be expressed in another domain. In this view, the gesture, story, or artifact that constitutes a specific creative act is not *what is evolving*; it is merely the external *manifestation* of the state of an evolving worldview. Like any evolutionary process, cultural evolution requires a balance between *novelty,* via the *generation* of variation, and *continuity,* via the *preservation* of variants that are adaptive. In cultural evolution, novelty is generated through creativity, and continuity is provided by social learning processes, e.g., imitation. Both the generative and imitative aspects of cultural evolution are affected by social media. We discuss the trajectory from social ideation to social innovation, focusing on the role of self-organization, renewal, and perspective-taking at the individual and social group level.

## Introduction

Sometimes it isn't until you take a step back from something and see it in context that you begin to understand it. When humans began to see Earth, not as the center of the universe, but as a planet revolving around our sun, a star, it catapulted us to a new level of understanding, and when we started seeing Earth from space as a pale blue dot we gained more perspective still. In this chapter we will take a step back from creativity and examine it in terms of the role it plays in fueling the evolution of culture.

When people interact with each other, directly through social exchange, or indirectly, by way of artifacts, we put our own spin on each others' ideas, adapting them to our own needs, tastes, and preferences. This not only generates cultural variation, it ensures that this variation is cumulative; my ideas build upon yours, and so forth, giving rise to a web of 'cultural lineages' extending through time and across space. The process is often exaptive: an object originally designed for one purpose may be redesigned to serve a different purpose (Gabora et al., 2013). So a first goal of this chapter is to take a step back from individual creative acts, and look at them as the novelty-generating component of an evolutionary process, the evolution of culture. A second goal of this chapter is to show how creativity and innovation can be applied to social situations to make them healthier and more effective.

## When Ideas Jump Ship: Cross-Domain Influence in Creativity

Let us start by looking at a fascinating phenomenon that arises in social creativity: the phenomenon wherein creative outputs in one domain inspire and influence creative outputs in another domain.

To see what makes this cultural phenomenon so intriguing, it is helpful to first compare it to its biological analog. In biological evolution, organisms must be members of the same species to mate, because they mate using instruction sets for how to build an organism like themselves—i.e., a self-assembly code—and their mate's self-assembly code must be compatible with theirs. (In biological organisms, self-assembly codes are made out of DNA, though in theory it could be made out of something else.) In cultural evolution, however, there is no self-assembly code; elements of culture rely on *us* to build them (Gabora, 2004, 2013). This means we can take two very different 'parent ideas', say, the idea of a beanbag, and the idea of a chair, and generate cultural novelty: a beanbag chair. Since neither a beanbag nor a chair replicates itself using a self-assembly code, the fact that they are very different did not thwart the beanbag chair invention process. This means that cultural lineages are highly subject to change.

In fact, ideas 'jump ship' regularly in all kinds of fascinating ways. For example, a study of cross-domain inspiration showed that it is possible to re-interpret a creative work in one medium into another medium. When painters were instructed to paint what a particular piece of music would 'look like' if it were a painting, naïve participants were able to correctly identify at significantly above chance which piece of music inspired which painting (Ranjan, Gabora, & O'Connor, 2013; Ranjan, 2014). Although the medium of expression was different, something of its essence remained sufficiently intact for people to detect a resemblance between the new creative output and its inspirational source. Thus, at their core, creative ideas are less domain-dependent than is widely assumed.

This conclusion received further support from another study, in which creative individuals in a variety of disciplines were asked to list as many influences on their creative work as they could (Gabora & Carbert, 2015). Of the 65 creative influences provided by the 66 participants, 47% were cross-domain influences (e.g., a painting influenced by music), 27% were narrow within-domain (e.g., a painting influenced by another painting), 8% broad within-domain (e.g., a painting influenced by sculpture), and 18% unclassifiable. This result surprised us, for we had just been looking to see if cross-domain influence exists at all; we were not expecting it to predominate! However, another study involving 261 undergraduate psychology students (i.e., unlike the previous study they were not particularly creative) yielded even more startling results: of the 508 creative influences listed, 67% were cross-domain, 13% were broad within-domain, 13% were broad within-domain, and 7% unclassifiable (Scotney, Weissmeyer, & Gabora, 2018).

Of course, evidence for the creative blending and contamination of ideas is everywhere, from fusion cuisine to Donald Duck slippers to 'Windows' computers. However, what these results show is that even when the creative *output* is not a blend but lies squarely in one domain, the *creative process giving rise to it* may be rooted in different domains. Thus, creativity is not confined to the particular 'problem domain' of the eventual creative output, as is widely assumed; creators can probe the vast hinterlands of their realities to scout out ingredients for their creativity.

### Creativity Fuels Cultural Evolution Through Cognitive Restructuring

At first glance it may seem that the basic units of cultural evolution would be such things as rituals or tools, but from the above evidence for the cross-fertilization of different domains, it seems the only way to delineate the 'conceptual parents' of a given idea is to look to the creator's entire web of knowledge and understandings. Not only could the hammer inspire another hammer design, it could inspire a song, or a Mickey Mouse cartoon. Thus, these findings set the stage for the framework for creativity proposed here, which is based on the argument that discrete elements of culture such as songs or stories are not what evolves through culture; they are the overt, observable manifestations of evolving cognitive structures (Gabora, 2004, 2008, 2013, 2017).

Indeed, it is widely thought that humans possess two levels of complex, adaptive, self-organizing, evolving structure: an organismic level and a psychological level (Barton, 1994; Combs, 1996; Freeman, 1991; Gabora, 1998, 2017; Pribram, 1994; Varela, Thompson, & Rosch, 1991). We can refer to this psychological level as a *worldview:* an individual's uniquely structured web of understanding that provides both a way of seeing the world and a way of being in the world, i.e., a mind as it is experienced from the inside. Thus, it is the worldview that is evolving through culture. It is the *self-organizing* nature of a worldview—i.e., the fact that it can continuously renew itself—that makes it impossible to trace all the influences or "conceptual parents" of a creative work such as a song or journal article. For example, consider the situation in which a video game inspires a song, which inspires a book. To see the thread of continuity across this "line of descent" it is necessary to consider how their creators navigate through webs of beliefs, attitudes, procedural and declarative knowledge, and habitual patterns of thought and action that emerge through the interaction between personality and experience.

The loosely integrated structure of a worldview enables us to detect gaps, inconsistencies, or problems, focus our attention on them, and reweave our conceptual webs to better understand or accept the situation, and find a solution, or at least a way of expressing our reaction. Honing an idea involves looking at it from the different angles proffered by one's particular worldview, 'putting ones' own spin on it', make sense of it in one's own terms, and expressing it outwardly (Gabora, 2017). It may involve the restructuring of representations by re-encoding the problem such that new elements are perceived to be relevant, or relaxing goal constraints (Weisberg, 1995). Thus, honing enables the creator's understanding of the problem or task to shift, and in so doing it may find a form that fits better with the worldview as a whole. In this way, not only does the task get completed (or worked on and put aside) but the worldview transforms, becomes more robust, and evolves. The transformative impact of immersion in the creative process extends far beyond the "problem domain"; it can bring about sweeping changes to that second (psychological) level of complex, adaptive structure that alter one's self-concept and view of the world. Creative acts and products render such cognitive transformation culturally transmissible. Thus, it is suggested that what evolves through culture is not creative contributions but worldviews, and cultural contributions give hints about the worldviews that generate them.

Although selection as the term is used in the layperson sense may play a role (people may be selective about which aspects of their worldviews they express, for example, or which paintings they show at a gallery), the cultural evolution process does not involve selection in its technical sense (change over generations due to the effect of differential selection on the distribution of heritable variation across a population). We posit that instead of search, creativity involves viewing the task from a new context, which may restructure the internal conception of it, and this restructuring may be amenable to external expression. This external change may in turn suggest a new context, and so forth recursively, until the task is complete.

A worldview not only self-organizes in response to perturbations but it is imperfectly reconstituted and passed down through culture. This is because it is not just self-organizing but *self-regenerating:* people share experiences, ideas, and attitudes with each other, thereby influencing the process by which other worldviews form and transform. Children expose elements of what was originally an adult's worldview to different experiences, different bodily constraints, and thereby forge unique internal models of the relationship between self and world. Thus, worldviews evolve by interleaving (1) internal interactions amongst their parts, and (2) external interactions with others. Through these social interactions, novelty accumulates and culture evolves. Elements of culture create niches for one another. One creative ideas begets another and modifications build on each other. This phenomenon, wherein there is an accretion of cumulative change over time, is sometimes referred to as the *ratchet effect* (Tomasello, Kruger, & Ratner, 1993).

### An Agent-Based Model of the Interplay of Creativity and Imitation in Cultural Evolution

EVOC (for EVOlution of Culture) consists of neural network based agents that invent new actions and imitate actions performed by neighbors (Gabora, 1995, 2008). The core of each agent is a simple neural network: an information processing unit inspired by how brains work, which can learn and generate ideas for different cultural outputs. For EVOC agents, the cultural outputs are actions, which get implemented as different combinations of movement across body parts. Although an agent is infinitely simpler than a human, its ideas (for actions) are integrated in the sense that they are encoded in overlapping distributions of 'neuron-like' interconnected nodes. Through the interplay of creatively building on existing ideas, and imitating what a neighbor is doing, the agents' cultural outputs evolve, i.e., exhibit cumulative, adaptive, open-ended change over time. Thus, the assemblage of ideas changes over time not because some replicate at the expense of others, as in natural selection, but through inventive and social processes. Agents can learn generalizations concerning what kinds of actions are useful, or have high 'fitness' with respect to a particular goal, and use this acquired knowledge to guide their creativity. A model such as EVOC is a vast simplification, and results obtained with it may or may not have direct bearing on complex human societies, but it allows us to vary one parameter while holding others constant and thereby test hypotheses that could otherwise not be tested. It provides new ways to think about and understand what is going on.

EVOC exhibits typical evolutionary patterns, such as (1) an increase in the fitness and complexity of cultural outputs over time, and (2) an increase in diversity as the space of possibilities is explored followed by a decrease as agents converge on the fittest possibilities, as illustrated in Fig. 1. It has been used to model how the mean fitness and diversity of cultural outputs is affected by factors such as population size and density, and borders between populations (Gabora, 1995, 2008), as well as the questions reported here pertaining to creativity.

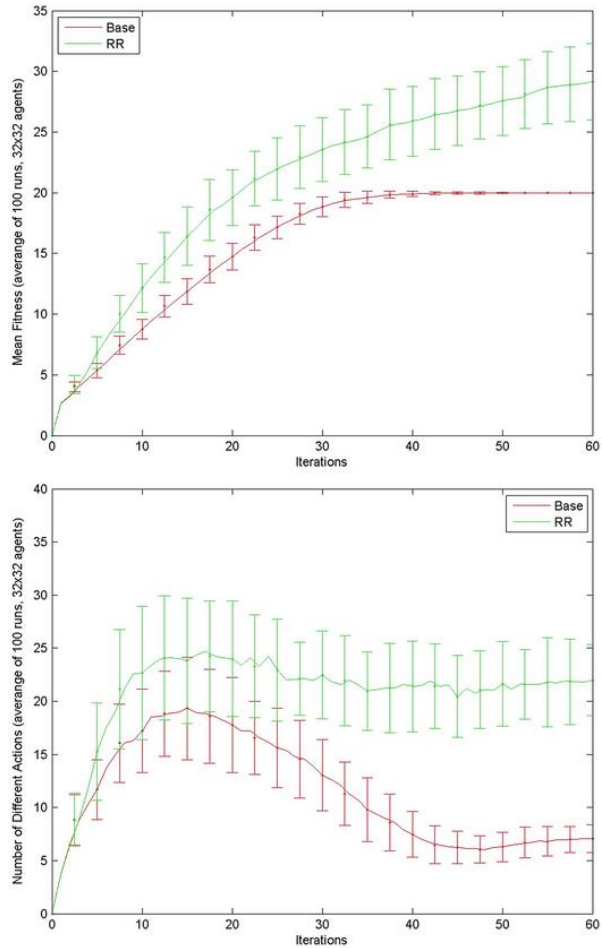

*Fig. 1:* *A typical graph of the increase in fitness of cultural outputs over time (top), and increase in diversity as the space of possibilities is being explored followed by a decline as the society converges on the fittest (bottom). These graphs also demonstrate the effect on fitness and diversity of a closed (labeled 'Base') versus open-ended (labeled 'RR', for 'representational redescription') space of possibilities.*

### Balancing Creativity with Continuity

Evolution—whether it be biological or cultural—combines processes that promote change with processes that promote continuity; in other words, it involves not just *generating* new possibilities, but *preserving* the best of them. It often also involves building on existing possibilities, which combines both generation and preservation. The point is: an evolutionary process is as much about holding onto stuff that works as coming up with new stuff. What this means for cultural evolution is that it isn't necessary for everyone to be creative for the benefits of creativity to be felt by all. Few of us can build a jet or sculpt a masterpiece, but they are nevertheless ours to use and enjoy. We can reap the rewards of a creative person's ideas by copying them, buying them, or just admiring them. When we kick back with a tub of Ben & Jerry's and binge-watch a season of Game of Thrones instead of working on a novel, it may not feel like we're contributing to cultural evolution, but we're absorbing all kinds of things about our culture—social norms and deviations from them, styles of fashion, diction, and so forth—that may in turn reveal themselves in our future interactions. It doesn't necessarily pay to be creative. While creative individuals generate the novelty that fuels cultural evolution, absorption in their creative process may impede the diffusion of proven solutions, effectively rupturing the fabric of society. This leads to the question: how creative should a society be? How much is too much?

    These are difficult questions to address in studies with real people, but it is possible to address them in EVOC. In a first experiment, all agents could both invent and imitate, and whether they invented or imitated on a given iteration was a probabilistic function of the invention-to-imitation ratio, which was varied systematically from 0 to 1 (Gabora, 1995). When agents never invented, there was nothing to imitate, so there was no cultural evolution at all. If the ratio of invention to imitation was just marginally greater than 0, not only was cumulative cultural evolution possible, but all agents eventually converged on optimal cultural outputs. When all agents only invented and never imitated, the mean fitness of cultural outputs was also sub-optimal because fit ideas were not diffusing through the artificial society, but cultural evolution took place nevertheless. Figure 2a and 2b show the impact over time of different ratios of inventing to imitating on the mean fitness and diversity, respectively, of cultural outputs across the artificial society. The society as a whole performed optimally with a mixture of inventing and imitating, the optimal ratio being approximately 1:1, with the exact value depending on the fitness function (i.e., the problem they had to solve); for example, with the difficult fitness function used to generate Fig. 2, it was significantly lower than 1:1. Unlike fitness, diversity of outputs was positively correlated with the ratio of creation to imitation, which makes sense, since creation resulted in new variants. These results supported the hypothesis that, as in biological evolution, culture evolves most effectively when novelty-generating processes (e.g., creativity) are tempered by continuity fostering processes (e.g., imitation).

    This finding that very high levels of creativity could be detrimental to society as a whole led to another hypothesis, which has to do with evidence compiled by (Florida, 2002) that individuals in a society naturally settle into two streams: the conventional workforce, and what he called the 'creative class'. We hypothesized that this division of labor has adaptive value for society as a whole. This was investigated in EVOC by dividing the artificial society into two types of agents: conformers that only obtained new actions by imitating, and creators that obtained new actions either by inventing or imitating neighbors (Gabora & Tseng, 2017). Each agent was either a creator or an imitator throughout the entire run, and whether a creator invented or imitated in a given iteration fluctuated probabilistically. We systematically varied C, the proportion of creators to imitators in the society, and p, how creative the creators were. As illustrated in Fig. 3, there was a tradeoff between C and p such that the more creators there were, the less creative they should be, and vice versa. This provided a different kind of evidence that society as a whole functions optimally when creativity is tempered with continuity.

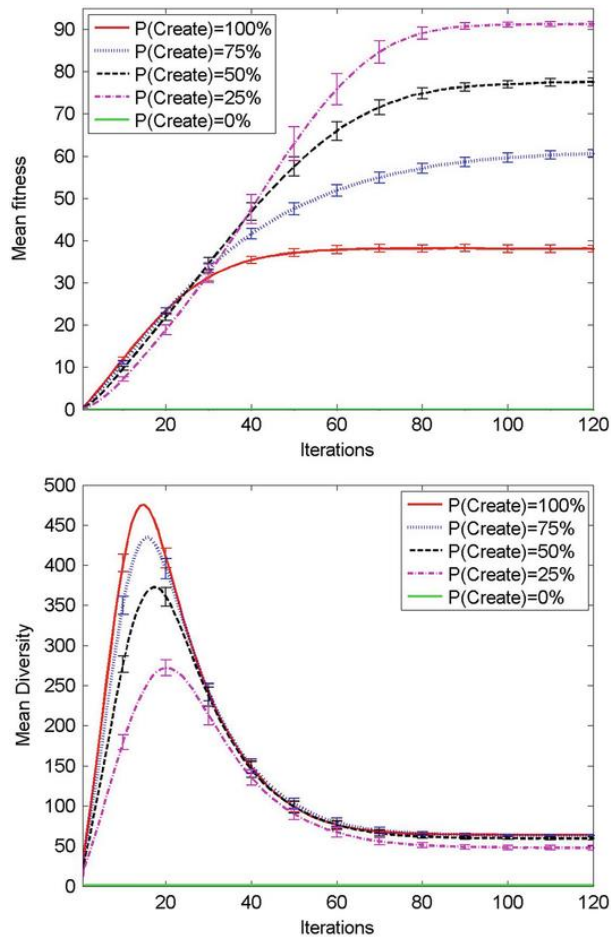

***Fig. 2:*** *Fitness (top) and diversity (bottom) of cultural outputs with different of invention to imitation ratios.*

This led us to hypothesize that society as a whole might benefit when individuals adjust how creative they are according to the perceived value of their creative outputs (possibly mediated by feedback from others). This too was investigated in EVOC (Gabora & Tseng, 2014). First we investigated whether giving agents the ability to self-regulate their creativity did indeed increase the mean fitness of ideas in the artificial society. Self-regulation (SR) of creativity refers to the capacity to modify how creative one is, potentially on the basis of external feedback from peers, but also potentially on the basis of hunches or intuitions about the potential of one's ideas. In EVOC, social regulation (SR) was implemented by enabling agents to increase their relative frequency of invention when they generated superior ideas, and decrease it when they generated inferior ideas. *p(C)* was initialized at 0.5 for both SR and non-SR societies. As illustrated in Fig. 4, with SR turned on, the mean fitness of the cultural outputs was higher than without it.

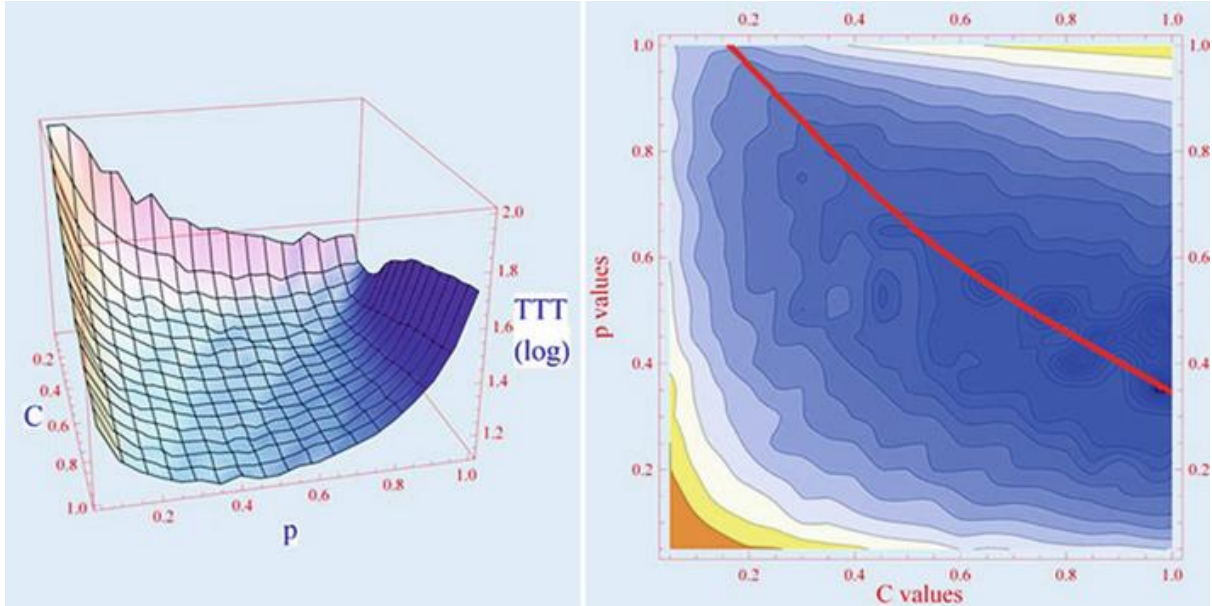

*Fig. 3.* The effect of varying the percentage of creators, C, and how creative they are, p, on mean fitness of ideas in EVOC. 3D graph (left) and contour plot (right) for the average mean fitness with different values of C and p. Since the z-axis is reversed to obtain an unobstructed view of surface, lower values indicate higher mean fitness. The red line on the contour plot indicates a clear ridge in fitness landscape indicating optimal values of C and p that are sub-maximal for most {C, p} settings, i.e., there is a tradeoff between how many creators there are and how creative they should be. (Adapted from Gabora & Firouzi, 2012.)

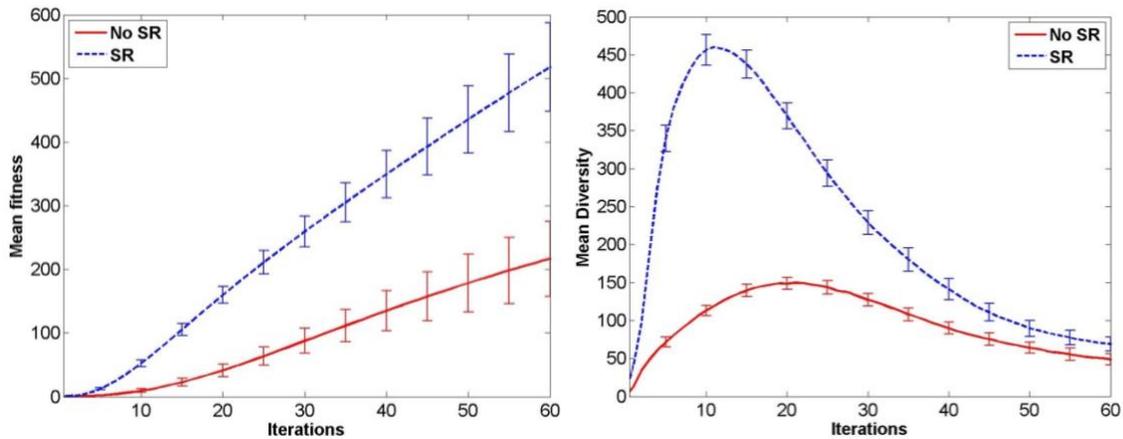

Fig. 4. shows the impact of SR on the mean fitness (left) and diversity (right) of cultural outputs. (Gabora & Tseng, 2014.)

Societies with SR segregated over time into two distinct groups: one that primarily invented, and one that primarily imitated. Thus, the increase in fitness could indeed be attributed to increasingly pronounced individual differences in their degree of creativity over the course of a run. Agents that generated superior cultural outputs had more opportunity to do so, and agents that generated inferior cultural outputs became more likely to use their iterations propagating other agents' ideas.

## Impact of Media on Creative Cultural Evolution

Another question that can be addressed with EVOC is: what is the impact of media on the cultural evolution of novelty? It's only been for the most recent sliver of human evolution that, by way of television, film, and radio, we've all had direct and immediate access to the same specific people. In one moment, all eyes can *converge* on Johnny Carson. Even more recently, with the internet, Netflix, and so forth, there has been a *divergence* of attention; at any time we can individually watch practically anything, or anyone, or even… Johnny Carson reruns.

The impact of media was investigated in EVOC using the *broadcasting* function (Gabora, 2008). Broadcasting allows a particular agent's actions to be accessed and imitated by not just immediate neighbors, but all agents. The broadcaster(s) can be selected or chosen at random before the run, or the user can specify that the agent with the fittest action is the broadcaster for that iteration. Adding one broadcaster produces a modest increase in the fitness of actions, but at the cost of greatly reduced diversity, since everyone starts doing what the broadcaster is doing. However, the more broadcasters there are, the less diversity is reduced.

We investigated the impact of broadcasting by comparing the diversity of actions in runs with zero, one, and five broadcasters, as illustrated in Fig. 5. In each case there was the usual increase in diversity (as the space of possibilities is explored) followed by a decrease (as agents converge on the best actions). However, with the addition of a broadcaster, the total number of different actions after 20 iterations decreased from eight to five, and the percentage of agents executing the most popular action increased from 41\% to 84\%. Thus, broadcasting accentuates the normal plummet in diversity. However, as we went from one broadcaster to five, the total number of different actions after 20 iterations increased from five to nine, and the percentage of agents executing the most popular action decreased from 84\% to 31\%. Thus, although media decreased diversity, this decrease in diversity was mitigated by more distributed media.

The effectiveness of creative versus uncreative broadcaster styles was also investigated in EVOC. Creative broadcasting increased the mean fitness of cultural outputs only when non-broadcasters were relatively uncreative, and increased the diversity of outputs only early in a run during initial exploration of the space of possibilities (Leijnen & Gabora, 2010).

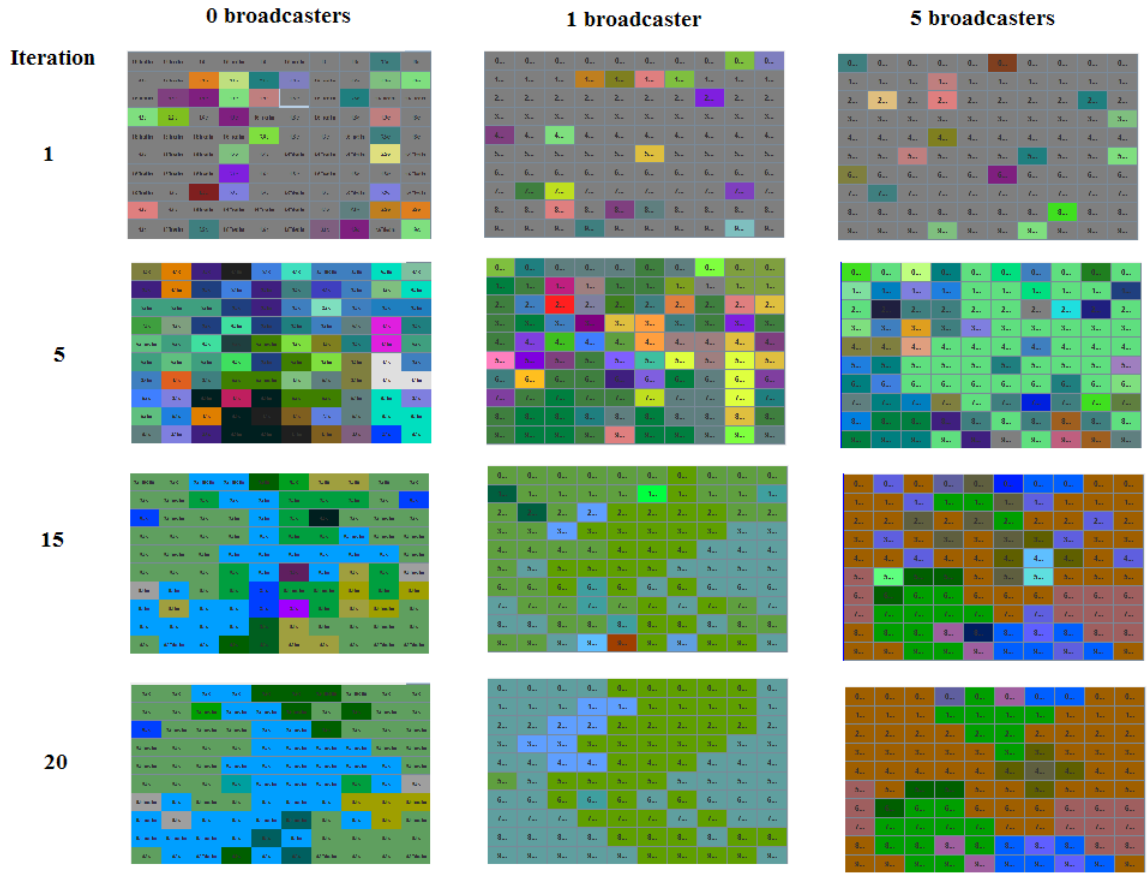

*Fig. 5. Diversity of actions after 1, 5, 15, and 20 iterations, over individual runs with 0, 1, and 5 broadcasters. Different actions are represented by differently colored cells. (From Gabora, 2008.)*

**Context, Perspective, and Point of View**

We said that honing an idea involves 'putting ones' own spin on it', make sense of it in their own terms, and expressing how it looks from one's point of view. It can be useful to differentiate between context, perspective, and point of view. A *context* can be defined as a specific lens that can affect the applicability of descriptors or the typicality of instances of a concept (Aerts & Gabora, 2005). For example, a guppy is not generally considered a typical FISH, but it is considered a typical PET FISH; in this case PET is acting as a context for FISH. A perspective can incorporate multiple contexts; for example, one perspective might take into account durability and utility while another takes into account cost of raw materials and value of final product (Veloz, Temkin, & Gabora, 2012). Perspective helps us organize our experiences, filter them through the conceptual schemes of our minds (Scheopner, 2013; Anderson, 1998). A point of view can be defined as a perspective that is associated with a particular individual, or individuals, owing to the situation(s) they are in, or 'where they stand' on one or more issues. Point of view is influenced by culture, as well as by needs, interests, and values. Thus, like a perspective, it can accommodate multiple contexts.

      An individual's point of view affects how he or she hones through an idea because it affects which contexts (i.e., which angles) the idea is considered from. Point of view can be

affected by mulling an idea over with others, for as Anderson (1998) notes, "the more eyes, different eyes, we can use to observe one thing, the more complete will our 'concept' of the thing, our 'objectivity,' be" (p. 18).

Context, perspective, and point of view are inherently social, in that they are affected by social interaction; this is the basis for the above-mentioned impact of positive impact of social imitation. This leads us to the topics of social creativity and social innovation.

## Social Creativity and Social Innovation

When humans began to see the Earth as a whole planet from space, we also began to contextualize our relationship to it in terms of how we impact it (Szerszynski & Urry, 2006). The social exchange we have with one another and the resulting cultural contributions that come out of those exchanges drive increasingly complex worldviews, and in turn an increasingly complex array of cultural outputs as new ideas build upon previous ideas. For example, my creative process culminates in a cultural product, which you come across on the internet, and it inspires you to create something similar through your own cultural lens. Someone else discovers your version of the product by accident through social media, and is stimulated to put their own spin on it, which leads to another product, and so on. Thus, the generative ideation process is highly social, in that social ties not only lead to novelty accumulation which evolves culture, but also pave the way for innovations that are socially inspired.

The term 'social creativity' refers to a type of creative process that involves one or more social 'layers'; in other words, it results from creative interactions amongst people, and can influence on social structures and networks. An example of social creativity might be a group of individual painters who decide to paint together in a social atmosphere and develop a particular 'style' that is considered innovative, influences other artists. Social creativity can bring about social change and culminate in complex, multilevel social systems (Moradkhah & Alborzi, 2016; Amabile, 1983; Domingues, 2000; Watson, 2007). To illustrate this using the above example, the painting group might receive escalating media attention, and start to take on social issues such as poverty in their art or discussions with media. It may be that not everyone agrees with their message, and the ensuing conflict encourages celebrities and politicians to get involved and support or reject the causes proposed, with their own creative adaptation or developments. *Innovation* is related to creativity, but the term innovation tends to be used when there is a distinct breadth and/or scale of *impact* to the creative endeavor, when the impact tends to remain *resilient* over time, and when the *implementation* of the endeavor has discernable methods or outcomes (Caulier-Grice, Patrick, & Norman, 2012). *Social Innovation* refers to the development of new products, services and programs when applied within social structures to meet the needs and transformations that occur within the cultural collective to improve social conditions (Hämäläinen & Heiskala, 2007). It also refers to the outcomes of social interactions between innovators (Mumford, 2002), and results in improvements to the quantity or quality of social life, defined less by novelty and more by consequences of implementation (Neumeier, 2012). Since social innovations arise to meet complex social problems, they often change the resources, routines, and belief structures of the system in which they arise (Westley & Antadze, 2010). A social innovation has resilience and broad impact, and thus is 'system-changing': it permanently alters the perceptions, behaviour and social structures that gave rise to these challenges. Simply put, a social innovation is an idea that works for the public good" (Caulier-Grice et al., 2012, p. 11).

Like social creativity, social innovation has multiple layers. It can entail reiterated creative interactions amongst individuals or collective networks possibly within complex multilevel social structures. It "transcends sectors, levels of analysis, and methods to discover the processes—the strategies, tactics, and theories of change—that produce lasting impact" (Phills, Deiglmeier, & Miller, 2008, p. 37). When a social innovation has durability and broad impact, it has aspects of *creative destruction*, which is economist Joseph Schumpeter's (1942) classic observation of how systems transform by persistently destroying old structures to create new ones. This means a social innovation can contain elements of *disruptive* and *catalytic innovation*, which are terms that refer to an innovative way of doing things that challenges the social structures we consider normal, and which may transform these structures or create new ones (Christensen, Baumann, Ruggles, & Sadtler, 2006). These new social structures may replicate previous successes; however, since they are scaled to match the current social need, they may generate financial, human or intellectual recourses that extend beyond the original system. As such, social innovations can "cross multiple social boundaries to reach more people and different people, more organizations and different organizations, organizations nested across scales (from local to regional to national to global) and linked in social networks" (Westley & Antadze, 2010, p. 4).

      Social innovations emerge out of the webs of understanding that constitute our worldviews, and they play a key role in cultural evolution. Social belief systems and habitual thinking patterns and behaviour can undergo profound transformation at a local, regional or global level when the impact of social innovations are far-reaching (Westley & Antadze, 2010). This suggests that collective worldviews can 'scale up' or change synchronistically across social networks due to a catalytic cross-fertilization or contamination of cultural domains. This is an indication of the self-regenerating quality of a worldview, and thus the adaptability and dexterity of creativity.

## Conclusion

Some say that having children makes us aware of being part of something infinitely larger than ourselves. Perhaps Mother Nature wired us up to feel this way because children are our lifeline into the future in terms of the propagation of our genes in biological evolution. When we're watching our child swim out to the island in the middle of the lake for the first time, nothing else in the world seems to matter or exist. But similarly, when we're in the throes of creative inspiration, it can feel like nothing else matters, like nothing else exists. Perhaps Mother Culture, the younger sister to Mother Nature but no less powerful, wired us up to feel this way because creative enterprises are our lifeline to into the future in terms of the propagation of our worldviews through a second evolutionary process: cultural evolution. Perhaps the time is ripe to heed the playful call of Mother Culture, to immerse ourselves in the thrill of creation and the satisfaction of innovation, as individuals and as groups, to reimagine and rebuild our relationship to this pale blue dot that we call home.

## Acknowledgements
This research was supported in part by a grant from the Natural Sciences and Engineering Research Council of Canada.

## References

Aerts, D., & Gabora, L. (2005). A state-context-property model of concepts and their combinations II: A Hilbert space representation. *Kybernetes, 34*(1&2), 192–221.
Amabile, T. M. (1983). *The social psychology of creativity*. New York: Springer-Verlag. https://doi.org/10.1007/978-1-4612-5533-8.

SOCIAL INNOVATION EVOLUTION WORLDVIEWSAnderson, R. (1998). Truth and objectivity in perspectivism. *Synthese, 115*(1), 1–32 Retrieved from http://www.jstor.org.ezproxy.library.ubc.ca/stable/20118040.

Barton, S. (1994). Chaos, self-organization, and psychology. *American Psychologist, 49*, 5–14.

Caulier-Grice, J., Davies, A., Patrick, R., & Norman, W. (The Young Foundation). (2012). Defining social innovation. *Social Innovation Overview: A deliverable of the project "The theoretical, empirical and policy foundations for building social innovation in Europe,"* (TEPSIE), European Commission – 7th Framework Programme, Brussels: European Commission, DG Research.

Christensen, C. M., Baumann, H., Ruggles, R., & Sadtler, T. M. (2006). Disruptive innovation for social change. *Harvard Business Review, 84*(12), 94–101.

Combs, A. (1996). *The radiance of being: Complexity, chaos and the evolution of consciousness*. St. Paul, MN: House Paragon.

Domingues, J. M. (2000). *Social creativity, collective subjectivity and contemporary modernity*. New York: St. Martin's.

Florida, R. (2002). *The rise of the creative class*. London: Basic Books.

Freeman, W. J. (1991). The physiology of perception. *Scientific American, 264*, 78–85.

Gabora, L. (1995). Meme and variations: A computer model of cultural evolution. In L. Nadel & D. Stein (Eds.), *1993 lectures in complex systems* (pp. 471–486). Boston: Addison-Wesley.

Gabora, L. (1998). Autocatalytic closure in a cognitive system: A tentative scenario for the origin of culture. *Psycholoquy, 9*.

Gabora, L. (2004). Ideas are not replicators but minds are. *Biology and Philosophy, 19*, 127–143.

Gabora, L. (2008). Modeling cultural dynamics. *Proceedings of the Association for the Advancement of Artificial Intelligence (AAAI) Fall Symposium 1: Adaptive Agents in a Cultural Context* (pp. 18–25). Menlo Park, CA: AAAI Press.

Gabora, L. (2013). An evolutionary framework for culture: Selectionism versus communal exchange. *Physics of Life Reviews, 10*, 117–167.

Gabora, L. (2017). Honing theory: A complex systems framework for creativity. *Nonlinear Dynamics, Psychology, and Life Sciences, 21*(1), 35–88.

Gabora, L., & Carbert, N. (2015). Cross-domain influences on creative innovation: Preliminary Investigations. In R. Dale, C. Jennings, P. Maglio, T. Matlock, D. Noelle, A. Warlaumont, & J. Yashimi (Eds.), *Proceedings of the 37th annual meeting of the Cognitive Science Society* (pp. 758–763). Austin, TX: Cognitive Science Society.

Gabora, L., & Firouzi, H. (2012). Society functions best with an intermediate level of creativity. In N. Miyake, D. Peebles, & R. P. Cooper (Eds.), *Proceedings of the 34th annual meeting of the cognitive science society* (pp. 1578–1583). Austin, TX: Cognitive Science Society.

Gabora, L., Scott, E., & Kauffman, S. (2013). A quantum model of exaptation: Incorporating potentiality into biological theory. *Progress in Biophysics & Molecular biology, 113*, 108–116.

Gabora, L., & Tseng, S. (2014). Computational evidence that self-regulation of creativity is good for society. In P. Bello, M. Guarini, M. McShane, & B. Scassellati (Eds.), *Proceedings of the 36th Annual Meeting of the Cognitive Science Society* (pp. 2240–2245). Austin, TX: Cognitive Science Society.

Hämäläinen, T. J., & Heiskala, R. (Eds.). (2007). *Social innovations, institutional change and economic performance: Making sense of structural adjustment processes in industrial sectors, regions and societies*. Cheltenham, UK: Edward Elgar Publishing.
13

Leijnen, S., & Gabora, L. (2010). An agent-based simulation of the effectiveness of creative leadership. In R. Camtrabone & S. Ohlsson (Eds.), *Proceedings of the Annual Meeting of the Cognitive Science Society* (pp. 955–960). Austin TX: Cognitive Science Society.

Mumford, M. D. (2002). Social innovation: Ten cases from Benjamin Franklin. *Creativity Research Journal, 14*(2), 253–266.

Neumeier, S. (2012). Why do social innovations in rural development matter and should they be considered more seriously in rural development research? – Proposal for a stronger focus on social innovations in rural development research. *Sociologia Ruralis, 52*(1), 50.

Phills, J. A., Deiglmeier, K., & Miller, D. T. (2008). Rediscovering social innovation. *Stanford Social Innovation Review, 6*(4), 34–43.

Pribram, K. H. (1994). *Origins: Brain and self-organization*. Hillsdale NJ: Lawrence Erlbaum.

Ranjan, A. (2014). *Understanding the creative process: Personal signatures and cross-domain interpretations of ideas*. Ph.D. Thesis, University of British Columbia, Canada.

Ranjan, A., Gabora, L., & O'Connor, B. (2013). The cross-domain re-interpretation of artistic ideas. Proceedings of the 35th Annual Meeting of the *Cognitive Science Society*(pp. 3251–3256). Houston, TX: Cognitive Science Society.

Sabelli, H., & Abouzeid, A. (2003). Definition and empirical characterization of creative processes. *Nonlinear Dynamics. Psychology, and Life Sciences, 7*, 35–47.

Scheopner, C. (2013). Perspectivism. In C. E. Cortés (Ed.), *Multicultural America: A multimedia encyclopedia* (Vol. 1, pp. 1695–1695). Thousand Oaks, CA: SAGE Publications Ltd. https://doi.org/10.4135/9781452276274.n686.

Schumpeter, J. (1942). *Capitalism, socialism and democracy*. New York: Harper.

Scotney, V., Weissmeyer, S., & Gabora, L. (2018). Cross-domain influences on creative processes and products. In (C. Kalish, M. Rau, J. Zhu and T. Rogers, Eds.) Proceedings of 40th Annual Meeting of the Cognitive Science Society. Austin TX: Cognitive Science Society.

Szerszynski, B., & Urry, J. (2006). Visuality, mobility and the cosmopolitan: Inhabiting the world from afar. *The British Journal of Sociology, 57*(1), 113–131. https://doi.org/10.1111/j.1468-4446.2006.00096.x.

Tomasello, M., Kruger, A. C., & Ratner, H. H. (1993). Cultural learning. *Behavioral and Brain Science, 16*, 495–511.

Varela, F., Thompson, E., & Rosch, E. (1991). *The embodied mind*. Cambridge MA: MIT Press.

Veloz, T., Temkin, I., & Gabora, L. (2012). A conceptual network-based approach to inferring the cultural evolutionary history of the Baltic psaltery. In N. Miyake, D. Peebles, & R. P. Cooper (Eds.), *Proceedings of the 34th Annual Meeting of the Cognitive Science Society* (pp. 2487–2492). Austin TX: Cognitive Science Society.

Watson, E. (2007). Who or what creates? A conceptual framework for social creativity. *Human Resource Development Review, 6*(4), 419-441. https://doi.org/10.1177/1534484307308255.

Weisberg, R. W. (1995). Prolegomena to theories of insight in problem solving: Definition of terms and a taxonomy of problems. In R. J. Sternberg & J. E. Davidson (Eds.), *The nature of insight* (pp. 157–196). Cambridge MA: MIT Press.

Westley, F., & Antadze, N. (2010). Making a difference: Strategies for scaling social innovation for greater impact. *The Innovation Journal: The Public Sector Innovation Journal, 15*(2), article 2. Retrieved from http://www.innovation.cc/scholarly-style/westley2antadze2make_difference_final.pdf